# Anisotropic light-electron-phonon coupling and ultrafast carrier separation in ferroelectric $BaTiO_3$

*A. B. Swain[1], S. Kale[2], R. Soni[2], P. Baum[1,*]*

[1] *Fachbereich Physik, Universität Konstanz, Konstanz, Germany*

[2] *Department of Physical Sciences, Indian Institute of Science Education and Research Berhampur, Berhampur, India*

E-mail: *peter.baum@uni-konstanz.de*

**Abstract**

**Ferroelectric materials with built-in electric fields are useful for ultrafast electronics and conversion of light into electrical energy, yet the interaction of carrier motion with ultrafast relaxation processes remains nontrivial. Combining ultrafast electron diffraction with electron microscopy of electromagnetic fields, we capture ultrafast lattice dynamics and nanometer-scale carrier transport in ferroelectric $BaTiO_3$. We discover that $BaTiO_3$ reacts to light with an anisotropic electron-phonon coupling that depends on the optical polarization of the excitation light. Excited electrons relax two times faster into phonons when the optical electric field aligns to the ferroelectric symmetry break. Furthermore, ultrafast electron electrometry captures the motion and separation of photo-excited electron-hole pairs in presence of the ferroelectric field. These combined results provide insight into the tangled and anisotropic interaction of photons with phonons and the ferroelectric field, producing phonons and voltages in a stepwise reaction path.**

## Introduction

The ultrafast and efficient conversion of light into other forms of energy is crucial for solar cells[1-3] high-speed electronics,[4,5] and devices based on quantum effects.[6,7] In ferroelectric materials,[5,8-16] a spontaneous symmetry break in the unit cell produces a static dielectric polarization and an intrinsic electric field of up to $2\times10^2$ MV $cm^{-1}$ at room temperature.[17-20] Such internal electric fields are useful for carrier separation in ferroelectric solar cells and can also serve for ultrafast electronic circuitry.[5,8,9] A key advantage of ferroelectric solar energy converter over silicon-based technologies is the ability of produce an above-bandgap photovoltage, enabling energy conversion beyond the Shockley–Queisser limit.[9,11,12,16]

When light is absorbed, it first creates excited electrons that in turn couple to the phonons and eventually heat up the material. At the same time, electron-hole pairs separate in the ferroelectric field and produce a photocurrent that can depend on the polarization of the incident light.[5, 10] Here, we investigate whether also the electron-phonon relaxation has such an effect. We unify time-resolved electron diffraction[21,22] with ultrafast electron microscopy of electromagnetic fields[23-26] to directly probe the coupled electronic and structural response of a $BaTiO_3$ thin-film to impulsive generation of electrons and holes. By tracking the phonon dynamics[21, 22] as a function of the incident polarization of the excitation light and the ultrafast deflection of the electron beam as a probe of time-frozen electric fields,[23] we become capable of disentangling the anisotropic reaction of the crystal lattice from the ultrafast carrier motion in the ferroelectric crystal field in space and time.

## Concept and experiment

Figure 1(a) depicts the perovskite-like unit cell of $BaTiO_3$ in the upward-poled configuration, where the Ba, Ti, and O atoms adopt an off-centered arrangement that generates upward-oriented Ti-O dipoles and induces a static polarization field. In the experiment (Fig. 1b), we apply femtosecond laser pulses (red) to excite $BaTiO_3$ (blue) with ferroelectric polarization $P$ along the $c$-axis (Fig. 1a) above the bandgap, analogously to the initial process in a photovoltaic device. Terahertz-compressed electron pulses[27] (blue) then probe the ultrafast structural dynamics of the atoms via electron diffraction[21,22] and the ultrafast photoexcited carrier motion via deflection of the electron beam.[23]

## Membrane fabrication and ferroelectric characterization

An ultrathin film of $BaTiO_3$ is grown by pulsed laser deposition on a 30-nm thin, mono-crystalline Si(100) support membrane of $100\times100\ \mu m^2$ in size. First, we grow a buffer layer of $SrRuO_3$ at 650°C at an oxygen partial pressure of 0.3 mbar, using an excimer laser with a fluence of 1.1 J $cm^{-2}$ and a repetition rate of 4 Hz. Subsequently, the $BaTiO_3$ layer to be investigated is grown in the same oxygen environment at 720 °C with a repetition rate of 2 Hz. The resulting thicknesses of $SrRuO_3$ and $BaTiO_3$ are ~7 nm and ~18 nm, respectively.

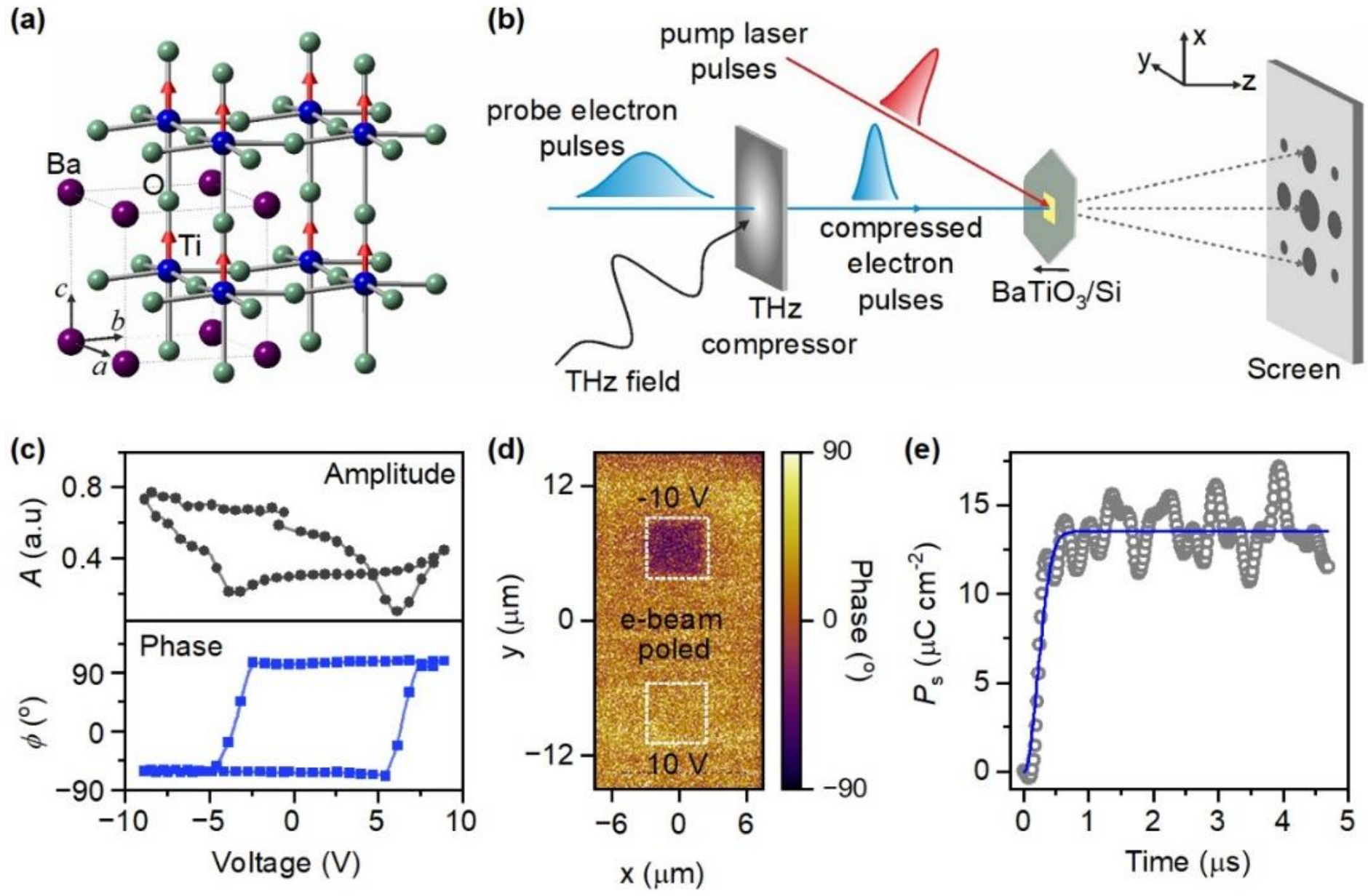

**Figure 1**: Material and experimental concept. (a) Crystal structure of tetragonal $BaTiO_3$, showing Ba atoms (violet), Ti atoms (blue), and O atoms (green) in off-centered positions along the *c*-axis. The upwards-displaced Ti atoms and resulting Ti-O dipoles generate a macroscopic ferroelectric polarization along the *c*-axis. (b) Schematics of the ultrafast electron diffraction and deflection experiment. Femtosecond laser pulses generate ultrashort electron pulses (blue beam) and terahertz pulses (black line) for compression of the electrons to ~70 fs in time[27]. The $BaTiO_3$ membrane on silicon (light yellow) is excited with laser pulses (red) and probed by time-delayed electrons (blue). The arrow denotes the ferroelectric field. (c) Measured amplitude $A$ and phase $\Phi$ of the ferroelectric hysteresis curve on the membrane. (d) Out-of-plane piezo force microscopy phase image, showing contrast within the electron-beam-poled region when applying a bias voltage of -10 V but not for +10 V (dashed regions). (e) Capacitive measurements (grey dots) show a stable ferroelectric polarization (blue) after impulsive electrical switch.

To determine the ferroelectric properties of our $BaTiO_3$ specimen, we apply piezo-response force microscopy (PFM). Figure 1(c) shows the measured local off-field hysteresis of the out-of-plane amplitude (upper panel) and phase (lower panel). For our ultrafast measurements, we need a 100 × 100 μm² region of poled material, more than can be produce with our piezo force microscope.[28] Therefore, we pole the as-grown $BaTiO_3$ membrane with a focused electron beam[29,30] in a scanning electron microscope at an electron beam energy of 30 keV. To verify the effective static poling by the electron beam, we overwrite the electron-beam-written spot (100 × 100 $\mu m^2$) with PFM-written spots (6 × 6 $\mu m^2$). Figure 1(d) show the results. We see in the background the electron-beam-written material and in the rectangles (white dashed lines) the overwritten domains. The sign of the measured pattern indicates that each incoming beam electron generates less than one secondary electron and therefore leaves a positive charge density on the surface that in turn polarizes the material.[29,30] We record the absolute polarization value using a positive-up-negative-down measurement by fabricating a capacitor around our $BaTiO_3$, where $SrRuO_3$ serves as the bottom electrode and a Pt layer as the top electrode. Figure 1(e) shows the measured polarization as a function of square pulse time. A fit[31] reveals a static polarization of $P_s = 12 \pm 2\ \mu C\ cm^{-2}$.

**Ultrafast structural dynamics**

In the ultrafast experiments, we pump this $BaTiO_3$ with laser pulses at a wavelength of 343 nm and photon energy of 3.6 eV, exceeding the bandgap (3.2 eV). The angle of incidence is 25° with respect to $BaTiO_3$'s $c$-axis, and the optical polarization is adjusted with a half-wave plate. The pump beam has a radius of ~160 µm, a pulse duration of ~200 fs, and a repetition rate of 40 kHz. The laser fluence is 0.45 mJ $cm^{-2}$, causing a long-time, thermodynamic temperature rise of ~60 K per pulse, calculated from the laser parameters, the measured absorption of our $BaTiO_3$ film (~69%), its heat capacity (0.5 kJ $kg^{-1}K^{-1}$)[32] and density (6.02 g $cm^{-3}$).[32] We therefore do not reach the ferroelectric Curie temperature of ~130 °C[32,33] and do not change the crystallographic phase.[21,22] The repetition rate is low enough to not produce an elevated pre-time-zero temperature.[34] Femtosecond electron pulses are generated by two-photon photoemission from a gold cathode and accelerated to 70 keV.[35] To minimize space-charge effects, we use only ~3 electrons per pulse.[36] Pulses are compressed to ~80 fs (full width at half maximum) by terahertz single-cycle pulses at a 10-nm thick aluminum membrane.[37] The electron beam diameter at the sample is ~80 µm, about two times smaller than the excitation beam. Data is measured on a phosphor screen with camera (F416, TVIPS).

Figure 2(a) shows the static electron diffraction pattern of our $BaTiO_3$ sample without laser excitation. For $BaTiO_3$, we observe all second-order spots {200}, the mixed-order spots $\{220\}$ and some of the fourth-order spots {400}. Features from diffraction rings from residual polycrystalline material are much weaker than the main Bragg reflections and do not affect the ultrafast analysis. From the silicon support membrane, we see all {220} and {400} spots. Faint additional spots are attributed to slight ferroelectric domain ordering or surface effects. Figure 2(b) shows a measured rocking curve of the $BaTiO_3$(200) reflection, revealing the expected $\sin(\theta)/\theta$ shape of our thin film, where $\theta$ is the angle of electron beam incidence. Figure 2(c) shows the schematic growth of $BaTiO_3$/Si heterostructure viewed along the [001] direction, where $BaTiO_3$(100) with $a = b \approx 3.99$ Å aligns to Si(100) with $a \approx 5.43$ Å at an angle of 45°.

We next measure (Fig. 2d) time-dependent Bragg spot intensities as a function of time and polarization state of the incoming light, to obtain insight into ultrafast electron-phonon coupling via the Debye-Waller effect.[38] Intensity traces are obtained by integrating around each spot with a radius of about three times the spot width and relative changes are evaluated by normalizing to pre-pump data at negative pump-probe delay. We measure 60 delay points, integrate for 5 s per image, and average over 75 delay scans.

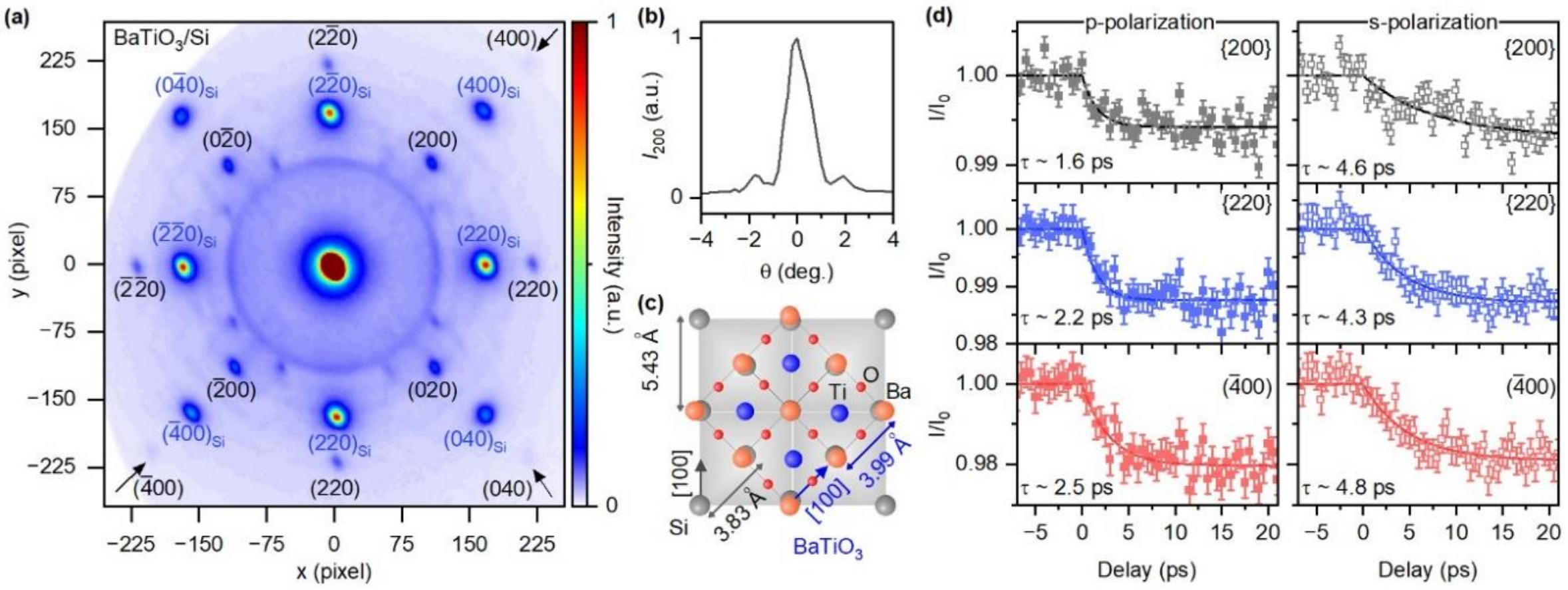

**Figure 2**: Ultrafast electron diffraction results. (a) Static electron diffraction pattern of the $BaTiO_3$/Si(100) heterostructure without laser excitation. Black labels indicate $BaTiO_3$ Bragg spots; blue labels indicate silicon spots. (b) Rocking curve of the $BaTiO_3$(200) spot. $I_{200}$, measured intensity; $\theta$, tilt angle. (c) Illustration of how $BaTiO_3$ aligns to silicon. Dark grey denotes Si, blue denotes Ti, orange denotes Ba, and red denotes O atoms. Light grey is the Si unit cell. (d) Measured ultrafast lattice dynamics of $BaTiO_3$ as a function of pump-probe delay. Left panel, time-resolved Bragg spot intensities $I/I_0$ for {200}, {220} and $(\bar{4}00)$ for excitation with p-polarized light. Right panel, measured Bragg spot intensities $I/I_0$ for excitation with s-polarized light. The solid lines are exponential fits.

Figure 2(d) shows the results for $I_{\{200\}} = \frac{1}{4}(I_{200} + I_{020} + I_{\bar{2}00} + I_{0\bar{2}0})$, $I_{\{220\}} = \frac{1}{4}(I_{220} + I_{2\bar{2}0} + I_{\bar{2}20} + I_{\bar{2}\,\bar{2}0})$, and $I_{(\bar{4}00)}$ of $BaTiO_3$. The left and right panels show the dynamics for p-polarization and s- polarization, respectively. A decrease of intensity is described by the Debye-Waller factor $\frac{I}{I_0} = exp\left[\frac{-1}{3}|G|^2 \cdot \langle u^2 \rangle\right]$, where $I$ is the measured intensity, $I_0$ is the intensity before the laser arrives, $G$ is the reciprocal lattice vector of the material and $\langle u^2 \rangle$ is the mean square displacement of the atoms. To quantify the speed and magnitude of this intensity drop, and thereby the electron-phonon coupling rate, we fit exponential functions according to $I_{hkl}(t) = 1 + \Delta I_{hkl}\left(e^{-(t-t_0)/\tau} - 1\right)H(t - t_0)$, where $I_{hkl}(t)$ is the intensity of $\{hkl\}$ as a function of time, $\tau$ is the rate constant, $H$ is the Heaviside function, $t_0$ is the laser arrival time and $\Delta I_{hkl}$ is the change amplitude. The solid lines in Fig. 2d show the results.

We see a substantial difference in the electron-phonon coupling rate that depends whether our material is excited with p-polarized (left panels) or s-polarized light (right panels). The rate for p-polarization is $\tau_p \approx$ 1.6-2.5 ps, like related materials,[39, 40] but the rate for s-polarization is $\tau_s \approx$ 4.3-4.8 ps, about two times slower. In contrast, the asymptotic intensity drops at >20 ps are identical for both cases; the {200}, {220}, and $(\bar{4}00)$ reflections show intensity drops of 0.5%, 1.2%, and 2.0%, in agreement with the expected scaling of the Debye-Waller effect with Bragg spot order. The total amount of energy that is deposited into the phonon bath is therefore independent of the laser's polarization state, that is, we reach the same final $\langle u^2 \rangle$, but the electron-phonon coupling rate differs by more than a factor of two.

These surprising results demonstrate that the vector orientation of our light wave controls the microscopic pathway of energy transfer from photons into electrons and phonons. Such an anisotropy can only be achieved if the laser-excited electrons in the band structure of $BaTiO_3$ not only inherit the optical polarization properties of the laser light but also transfer their properties to the phonons before electron-electron scattering or related processes can randomize the electrons. Such an effect is unexpected because electron-electrons scattering is usually much faster (few femtoseconds) than electron-phonon coupling (several picoseconds). In addition, we apply in the experiment a laser incidence angle of 25° and our p-polarized laser light, therefore, has only $\sin(25°) \approx 40\%$ field strength or $\sin^2(25°) \approx 18\%$ intensity along the *c*-axis and still $\cos(25°) \approx 90\%$ field strength or $\cos^2(25°) \approx 82\%$ intensity in the *a*-*b*-plane. If *a*,*b*-axis electronic states and *c*-axis electronic states would individually relax into the measured phonons with different rates, we would expect a double-exponential decay in the experiments, contrary to the single-exponential functions that are observed (Fig. 2d). Compared to the s-polarized excitation with 100% field strength and intensity in the *a*-*b*-plane (Fig. 2d, right side), the p-polarized experiment (Fig. 2d, left side) adds only a small fraction of c-polarized optical fields. Nevertheless, we see a catalysis-like speed-up by a factor of more than two of the overall electron-phonon relaxation rate.

To elucidate possible mechanisms, we invoke the real-space atomic motions in symmetry-broken $BaTiO_3$. The local potential energy surface of Ti atoms moving in the ferroelectric, tetragonal unit cell is given by $V = \frac{1}{2}\alpha \sum_i u_i^2 + \frac{1}{4}\lambda \sum_i u_i^4 + \gamma \sum_{i,j\,(i\neq j)} u_i^2 u_j^2$,[41-44] where $u_i$ is the local atomic displacements along $i = a, b, c$ directions, respectively. $\alpha$ is the harmonic coefficient, $\lambda$ is the anharmonic coefficient, and $\gamma$ is the coupling coefficient. In this double-well potential,[41-44] the minima correspond to the symmetry-broken equilibrium position of Ti shown in Fig. 3. Around this equilibrium, we obtain $V_{c,ab} = \frac{1}{2}\alpha u_c^2 + \frac{1}{4}\lambda u_c^4 + \frac{1}{2}\beta u_{ab}^2 + \gamma u_c^2 u_{ab}^2$, where $\beta$ is the harmonic coefficient, $u_{ab}$ is displacement in the *a*-*b* plane, and $u_c$ is the *c*-axis displacement. While the $u_c$ and $u_{ab}$ terms define the basic potential wells along *a*,*b* and *c*, the $\gamma u_a^2 u_{ab}^2$ describes non-equilibrium energy flow. Figure 3 sketches in the lower part this double-well potential, in which symmetry is broken along the *c*-axis (left-lower part) but oscillation is mostly harmonic in the *a*,*b*-plane (right-lower part). If we apply s-polarized laser light, we excite Ti $3d_{x,y}$-bands along the optical polarization axis. The measured 4.5-ps time scale (Fig. 2d, right panels) shows that these electronic states relax rather slowly into phonons (green) in the *a*-*b*-plane, the ones measured in the diffraction experiment. In contrast, if we apply p-polarized laser light, we also (in addition; see above) excite a certain fraction of electrons of Ti $3d_z$ band. If we assume that these electronic states can rapidly and very efficiently couple to the soft *c*-axis phonons of ferroelectric $BaTiO_3$,[45] we will rapidly create a broadened *c*-axis phonon state (green, dashed). Anharmonic coupling[46] then softens or hardens the *a,b*-axis phonons (dashes black line). If so, we may thus open up new channels (dashed lines) for relaxation of the (still remaining) excitations in the $3d_{x,y}$ bands (from the other component of our p-polarized laser light at 25°) that now have relaxed conditions for momentum transfer. In this picture, the rapid population of *c*-axis phonons (not directly measured) thus catalyzes the overall electronic relaxation into the *a-b*-plane phonons (measured). In addition, it is also possible that electronic effects[10-12] or changes of soft-mode susceptibility with laser frequency[47] play some role.

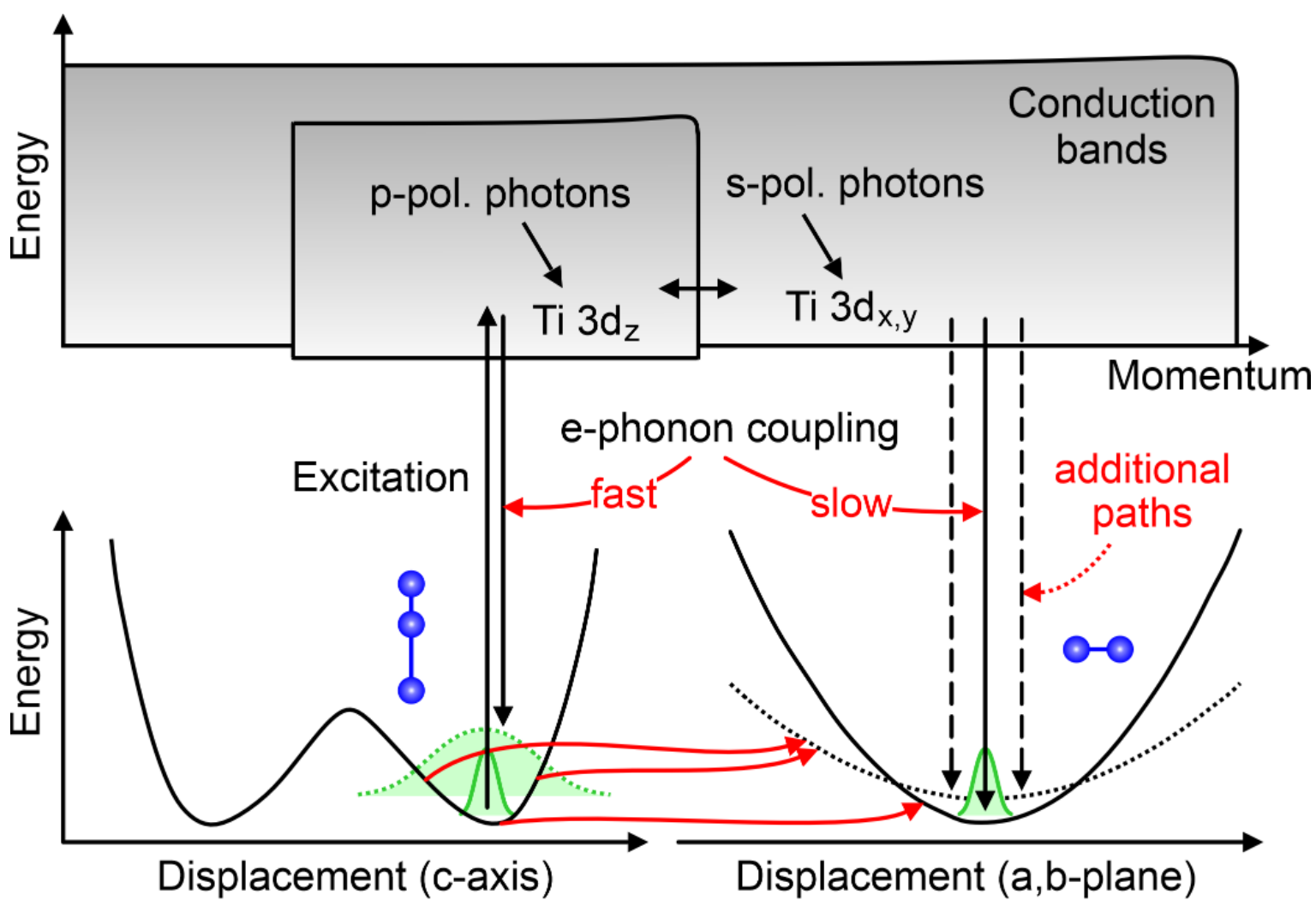

**Figure 3**: Potential mechanism for anisotropic light-electron-phonon coupling in $BaTiO_3$. Upper part, sketch of the conduction bands (light grey) formed by Ti ions. Lower part, cuts through the free energy surface of the phonons along the *c*-axis (left) and *a,b*-axis (right) as a function of Ti-displacement from its symmetry-broken equilibrium position. Blue dots, simplified sketch of the atomic positions in symmetry-broken (left) and symmetric (right) geometries (compare Fig. 1a). Green, vibrational ground state. Excitation with s-polarized photons (with electric field in the *a-b*-plane) creates $3d_{x,y}$-band electrons in the conduction band (light grey). These states relax slowly (~4.5 ps) to phonons in the *a-b*-plane (right back downwards arrow). Excitation with p-polarized photons (with electric field components along the *c*-axis) creates $3d_z$-band electrons (left black upwards arrow). These states rapidly create *c*-axis phonons (green dashed line) that soften the *a-b*-plane curvature (dotted black line). Conduction band electrons from $3d_{x,y}$-bands now have additional relaxation paths (dashed lines). Laser fields along the *c*-axis can therefore catalyze and speed up the relaxation of *a-b*-plane electrons (~2 ps).

**Ultrafast electron electrometry**

In a second experiment (Fig. 4a), we use ultrafast electrometry to measure and quantify the ultrafast diffusive carrier separation in our ferroelectric material. We let the electron beam (blue) hit the material (grey) under an angle of 10-45°. Initially (upper panel), the ferroelectric polarization (black arrows) is non-normal to the electron beam, causing a static beam deflection. After laser excitation (bottom panel), the deflection is reduced as a function of delay time due to time-frozen Lorentz forces of moving charges (+ and -) in the material.[23] We expect a deflection by an angle $\alpha_{x,y}\,(x,y,t) \approx \frac{qd}{m_e v^2} E_{x,y}(t)$, where $E_{x,y}(t)$ is the instantaneous electric field in $BaTiO_3$, $q$ is the charge of the electron, $d$ is the thickness of the specimen, $m_e$ is the electron mass and $v$ is the electron velocity.[23] The electron is here assumed to be fast enough to transverse the sample in sub-femtosecond time[48] and only obtain an angle, not a curved trajectory.[23] We see that the measured beam deflection angle is directly proportional to the photoinduced local and instantaneous electric field inside of the material.[26]

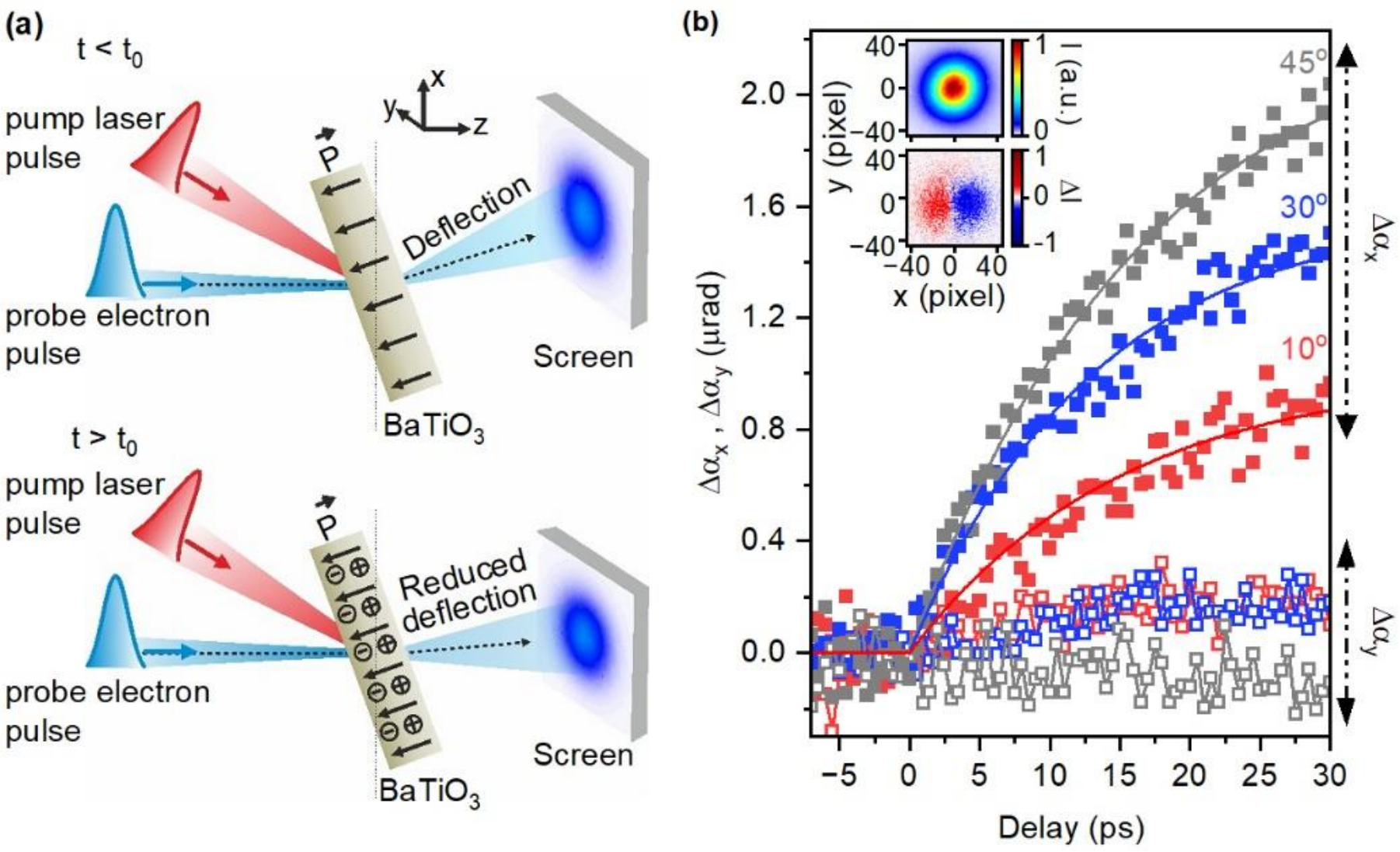

**Figure 4**: Ultrafast electron electrometry of photoexcited electron-hole motion in ferroelectric $BaTiO_3$. (a) Schematic of the pump-probe geometry. Before laser arrival (upper panel), the intrinsic ferroelectric field (black arrows) in poled $BaTiO_3$ deflects the electron beam (blue) in a static way (dotted line). After laser excitation (lower panel), the photoexcited electron-hole pairs start to separate and screen the effective electric field (black arrows). The electron beam obtains a reduced deflection angle (dotted line) as a function of time delay. (b) Measured beam deflection along x (squares) and y (open squares) for specimen angles of 10° (red), 30° (blue) and 45° (grey). The upper insert shows the measured electron beam profile $I$ before laser incidence (top panel) and the lower inset the measured change $\Delta I$ of the direct electron beam profile at a delay of 15 ps.

In this experiment, we apply the same laser fluence of 0.45 mJ $cm^{-2}$ but we now tilt the specimen at an angle of 10°, 30° and 45° with respect to the incoming electron beam. The insets of Fig. 4b show the profile of the incoming electron beam (top panel) and its change after 10-20 ps (bottom panel). The observed bipolar contrast (red and blue) is characteristic of a lateral translation of the central beam along $x$ but not along $y$ on the camera at no substantial change of beam shape or width.

Relative deflection angles $\Delta\alpha_x$ and $\Delta\alpha_y$ as functions of time are obtained from the camera distance (1.34 m) by Gaussian fits of the beam profile with respect to the position of the pre-pump beam. Figure 4(b) shows the measured deflections $\Delta\alpha_x$ and $\Delta\alpha_y$ as a function of pump-probe delay for angles of 10°, 30° and 45°. The solid squares show the deflection along $x$ and the open squares refer to deflections along $y$. The deflections along $x$ increase with angle and the deflections along $y$ are approximately zero.

Residual signals may originate from small magnetic effects.[23] We fit the experimental data in time with an exponential function $\Delta\alpha_x(t) = \Delta\alpha_x(0)e^{-(t-t_0)/\tau_{sep}}H(t-t_0)$, where $\Delta\alpha_x(t)$ is the measured deflection angle, $\Delta\alpha_x(0)$ is the deflection amplitude, $\tau_{sep}$ is the rate constant, $H$ is the Heaviside function, and $t_0$ is the laser arrival time. The solid lines in Fig. 4b show these fit results. The deflection at 45° has a time constant of 13 ps and an absolute deflection angle of $\Delta\alpha_x \approx 2.1\ \mu rad$. Similarly, the deflection at 30° and 10° has a time constant of 13 ps and 15 ps with an absolute deflection angle of $\Delta\alpha_x \approx 1.4\ \mu rad$ and $0.8\ \mu rad$, respectively. We do not find evidence of a polarization-dependent transport.

This beam deflection reflects the change of the effective electric field inside $BaTiO_3$ by electron-hole separation in the ferroelectric field. We expect a depolarization field of $E_{dp} \approx \frac{-P_s}{\varepsilon_r \varepsilon_0}$,[49, 50] where $P_s$ is the measured ferroelectric polarization strength ($12\ \mu C\ cm^{-2}$) and $\varepsilon_r \approx 100$ is the dielectric constant of $BaTiO_3$.[51] We obtain $E_{dp} \approx$ -1.3 MV $cm^{-1}$ along the *c*-axis. From the laser fluence of 0.45 mJ $cm^{-2}$, the optical beam radius of ~160 μm, the measured optical absorption at the membrane of ~68% and the film thickness of ~18 nm, we estimate a photoexcited electron-hole pair density of $\Delta N \approx 10^{20}$ $cm^{-3}$. The corresponding surface change density associated with polarization screening is $\sigma_{eh} = f\Delta Nqd \approx 2.8\ \mu C\ cm^{-2}$, where $f \approx 0.1$ represents the fraction of photoexcited carriers that contribute to screen the polarization field[50] while others recombine or become trapped. The resulting maximum photo-induced change to the effective electric field is therefore $E_{ind} = \frac{\sigma_{eh}}{\varepsilon_0 \varepsilon_r} \approx 0.33$ MV $cm^{-1}$. The estimated deflection amplitude along the *x*-axis becomes $\Delta\alpha_x \approx \frac{qd}{m_e v^2}E_{ind}(t)\sin\theta \approx 3.5\ \mu rad$ at $\theta = 45°$, in reasonable agreement with the experimentally observed result (~2 $\mu rad$).

From the measured time constants, we can determine the carrier mobility $\mu$ in $BaTiO_3$. Using Maxwell's equation $\nabla \times H = J + \frac{dD}{dt}$ and the current relation $J = qf\Delta N\mu E_{ind}(t)$, we find that the time-dependent electric field inside our material satisfies $\frac{dE}{dt} + \left(\frac{qf\Delta N\mu}{\varepsilon_0\varepsilon_r}\right)E = 0$. Solving this differential equation yields an exponential decay of $E_{ind}(t) = E_0 e^{-t/\tau_{sep}}$, where the decay constant $\tau_{sep} = \frac{\varepsilon_0\varepsilon_r}{qf\Delta N\mu}$, and $\mu = \frac{\varepsilon_0\varepsilon_r}{qf\Delta N\tau_{sep}}$. Using our observed average value for $\tau_{sep} \approx 14$ ps, we obtain a mobility of $\mu \approx 0.39$ $cm^2\ V^{-1}s^{-1}$ at ~80 °C, after femtosecond heating (see above). This value agrees with steady-state measurements,[52,53] showing that ultrafast carrier dynamics in $BaTiO_3$ on picosecond time scales and nanometer length scales is mostly diffusive.

**Conclusion and outlook**

From the combined results we conclude that photo-excited electron-hole pairs in $BaTiO_3$ relax into phonons in three distinct ways. First, the optical energy of the p-polarized component of the optical electric field, if present, is quickly transferred to in-plane phonons along the *a-b*-axis within less than 2.5 ps. Second, for s-polarized light without an electric field component along *c*-axis, *a-b*-plane phonons excited at substantially slower rates of only 3.7-4.8 ps (right panel in Fig. 2d). Third, on a much slower time scale of ~14 ps, after thermalizing of the phonon bath, the photoexcited and relaxed electron-hole pairs start to separate and build up a macroscopic internal electric field in opposite direction to the polarization field. The clear separation of these timescales and mechanisms demonstrates that energy relaxation from the optical field to the electrons to the phonons proceeds stepwise in an anisotropic way and precedes carrier transport and polarization screening. If carrier separation were faster than electron–phonon relaxation, transport would involve hot carriers, which is not observed.

In terms of technology, the reported combination of ultrafast electron diffraction and ultrafast electron-beam electrometry provides a unified, direct access to the interplay between the electric field of a light wave, the generation of electrons and holes, the subsequent electron-phonon coupling into different crystal directions, and the diffusive motion of electrons and holes by transport in a single experiment. We achieve a direct, contact-free measurement of mobility under true non-equilibrium conditions on picosecond timescales, without the need for electrical contacts, electrodes, or steady-state assumptions of conventional field-effect transistor measurements.[54, 55] In contrast to optical pump-probe reflectivity or terahertz conductivity methods,[56, 57] which infer mobility indirectly through model-dependent fitting of conductivity spectra, our novel ultrafast electron deflection measures the actual dynamics of the space-charge field in thin materials directly in space and time. We argue that such a combined ultrafast electron metrology, that captures electromagnetic phenomena and atomic motions on equal grounds, will become a valuable tool to explore the dynamics of simple and complex materials. There is potential for unexpected discoveries, here the finding of an anisotropy of the ultrafast response in ferroelectric materials to differently polarized laser light.

## Methods

### Thin film growth

The $BaTiO_3$ membrane is fabricated using pulsed laser deposition on a 3-mm diameter Si(100) grid (Silson), having 3×3 suspended membrane windows of 100×100 μm$^2$ size. The membrane thickness is 30 nm. First, the silicon grid is ultrasonically cleaned using acetone and isopropyl alcohol. The native oxide layer is removed by dip in a 3% hydrogen fluoride solution for 25 s, followed by drying with dry nitrogen and bath in methanol and isopropyl alcohol. The structure is then loaded to a resistive-type heater and transferred immediately to the deposition chamber, pumped down to a pressure of $5\times10^{-8}$ mbar. For pulsed laser deposition, we use a KrF excimer laser with a wavelength of 248 nm. A $SrRuO_3$ buffer layer is grown at 650°C in an oxygen partial pressure of 0.3 mbar, with a laser fluence of 1.1 J cm$^{-2}$ and a repetition rate of 4 Hz. To prevent native oxide formation due to residual oxygen, the first 100 pulses of the $SrRuO_3$ layer are deposited in a vacuum environment without oxygen. After a total of 1800 pulses, the $SrRuO_3$ thickness is estimated to be ~8 nm. Subsequently, $BaTiO_3$ is deposited on the $SrRuO_3$ layer without altering the deposition environment at 720°C with a laser fluence of 1.1 J cm$^{-2}$ and a repetition rate of 2 Hz. A total of 2300 pulses were used, corresponding to a thickness of ~18 nm, with a growth rate of ~50 pulses per unit cell. The total bilayer thickness is thus ~25 nm.

**Ferroelectric characterizations**

To investigate the ferroelectric properties of the grown $BaTiO_3$, we apply piezoresponse force microscopy, PFM (Asylum Research, Oxford Instruments). The switching spectroscopy channel is used to probe the local ferroelectric behavior by analyzing the phase and amplitude hysteresis loops, and the out-of-plane PFM imaging technique is used in dual resonance tracking mode to visualize the phase dynamics of the written ferroelectric domains. To confirm the areal ferroelectricity, a voltage bias greater than the switching threshold is applied to the scanning tip, which acts as the top electrode, while the bottom $SrRuO_3$ electrode remains grounded. To measure the resulting polarization switch, we apply a tip bias of ±10 V (see Fig. 1d). The resulting absolute domain orientation and poling direction is then confirmed by out-of-plane PFM imaging. Quantitative polarization measurements are performed by fabricating capacitor structures. Platinum top electrodes (5 µm diameter) are deposited through a stencil mask. Transient switching current measurements are conducted using a five-pulse sequence (5 µs pulse width, 4 V amplitude). The intrinsic switching current is isolated by subtracting the non-switching (background) transient from the switching response, following the methodology of Ref. 31. The time-dependent switched polarization and the static polarization values are extracted by fitting the data to the Kolmogorov-Avrami-Ishibashi (KAI) model.[31]

**Pump-probe ultrafast electron diffraction**

To probe the structural dynamics and the transient electric fields of the ferroelectric $BaTiO_3$ membrane, we employ a pump-probe ultrafast electron diffraction setup in a transmission geometry.[48] For this pump-probe studies, ~250fs laser pulses used for the sample excitation at a repetition rate of 40 kHz are obtained from an Yb:YAG laser system (Pharos, Light Conversion UAB). Knife-edge measurements reveal an optical beam radius 200 µm at an intensity of $1/e^2$ at the sample position. The excitation fluence at the beam center is set to ~0.45 mJ $cm^{-2}$. For ultrafast electron diffraction, electron pulses are generated by two-photon photoemission on a gold cathode and accelerated to a kinetic energy of 70 keV.[35] The electron beam diameter (full width at half maximum) at the specimen is ~80 µm. The electron pulses are compressed in time by single-cycle THz pulses at an ultrathin mirror membrane of 10 nm Al on a 10 nm SiN support membrane,[37] placed at 0.4 m from the electron source and 0.5 m before the specimen. The detector is placed 1.4 m after the specimen. Terahertz streaking is used to characterize the electron pulse duration at the specimen location,[27] using a bow-tie resonator (copper, 400 µm width, 400 µm height, 60 µm gap width, 80 µm gap height). To improve the signal, we insert a 50-µm aperture into the beam before the resonator. The measured electron pulse duration is 80 fs. Time-resolved electron diffraction images are accumulated over ~75 pump–probe time scans per experiment (Fig. 2d. For the electron electrometry measurements, we record center-of-mass shift of the direct electron beam as a function of the pump-probe delay. To ensure statistical significance, both the diffraction and deflection data are integrated over multiple (typically ~75) pump-probe cycles in each experiment (Fig. 2d, Fig. 4).

**Acknowledgments**

This project has received funding from the Deutsche Forschungsgemeinschaft (DFG) via SFB 1432, the European Research Council (ERC) through AdG ULMI, and the Marie-Skłodowska-Curie Grant 101064961-SpaceTimeFerro.

**Author contributions**

P.B and A.B.S conceived the concept and the experiment. A.B.S performed the diffraction and electrometry experiment and analyzed the data, S.K and A.B.S produced the specimen under the supervision of R.S. S.K and R.S performed the ferroelectric characterizations, P.B and A.B.S wrote the manuscript with the help of all co-authors.

**The authors declare that they have no competing financial interests.**

**Data is available from the authors on request.**